\documentclass[a4paper,showpacs,twocolumn,amsmath,amssymb,floatfix,prl,preprintnumbers,footinbib]{revtex4}
\usepackage{graphicx}
\usepackage{dcolumn}% Align table columns on decimal point
\usepackage{bm}
\usepackage{nicefrac}
\usepackage{mathbbol}
\usepackage{dsfont}
\usepackage{color}
\usepackage{url}
\usepackage[colorlinks=true,citecolor=blue]{hyperref}
\usepackage{amsmath,amssymb}
\catcode`\"=\active \let"=\" \let\3=\ss
\newcommand{\bra}[1]{\langle #1|}
\newcommand{\ket}[1]{|#1\rangle}
\newcommand{\vf}{v_{\mbox{\tiny F}}}
\begin{document}
\title{Flat bands and long range Coulomb interactions:
conducting or insulating~?}
\author{Wolfgang H\"ausler}
\affiliation{Institut f\"ur Physik, Universit\"at Augsburg,
D-86135 Augsburg \ and\\
I.~Institut f\"ur Theoretische Physik, Universit\"at Hamburg,
D-20355 Hamburg, Germany}
\begin{abstract}
Dispersionless (flat) electronic bands are investigated
regarding their conductance properties. Due to ``caging'' of
carriers these bands are usually insulating at partial filling,
at least on the non-interacting level. Considering the specific
example of a $\mathcal{T}_3$--lattice we study long-range
Coulomb interactions. A non-trivial dependence of the
conductivity on flat band filling is obtained, exhibiting an
infinite number of zeros. Near these zeros, the conductivity
rises linearly with carrier density. At densities half way
in between adjacent conductivity-zeros, strongly enhanced
conductivity is predicted, accompanying a solid-solid phase
transition.
\end{abstract}
\pacs{71.10.Fd,71.10.-w,73.20.Qt,73.20.Jc,73.90.+f}
\maketitle

Recently, electronic flat bands in periodic lattices have
received considerable attention \cite{reviews}, where single
particle energies $\varepsilon_k$ of at least one tight binding
band stay non- or weakly dispersing, throughout the Brillouin
zone. One focus of interest in two spatial dimensions, similar
to Landau levels, is the effect of interactions which in the
absence of kinetic energy always has to be treated
nonperturbatively. Fractional Chern insulator (FCI) phases have
been identified \cite{neupert11,daghofer12}, some of which
exhibit Chern numbers larger than unity \cite{wang11} and
thereby generalize fractional quantum Hall states. Meanwhile,
many lattices have been detected to host flat bands. Band
flatness \cite{nomass} arises due to localization by local
quantum interferences, coined \cite{doucot98} as ``caging'' of
carriers. As a result, the conductivity vanishes, at least on
the noninteracting level. This accords with the vanishing Chern
number of {\em strictly\/} flat single particle bands where
$\:{\rm d}\varepsilon_k/{\rm d}k=0\:$, throughout the Brillouin
zone, as proven recently \cite{chen14} for tight binding
lattices.

On-site Hubbard interaction seems to delocalize vicinally caged
carriers, which was considered as indication for nonzero
conductivity \cite{doucot00}. However, short range interactions
cannot impair the huge flat band degeneracy at low fillings
\cite{dassarma09}. Competing charge density wave phases have
been studied \cite{daghofer12}. Here, we investigate long range
Coulomb interactions which at any filling will lift the flat
band degeneracy. In quantum Hall systems, when kinetic energy
is quenched, they cause Wigner crystallization at fillings
$\nu<\nicefrac{1}{5}$ \cite{stormer90} (in graphene at
$\nu<0.28$ \cite{moraissmith09}) while without magnetic field
crystallization appears when the Coulomb energy $E_{\rm C}$
exceeds the kinetic energy $E_{\rm K}$ by a sufficiently big
factor \cite{tanatar}, $\:E_{\rm C}/E_{\rm K}>37\:$. The
longitudinal conductivity
\begin{equation}\label{sigmaomega}
\sigma(\omega)=D\:\frac{1}{{\rm i}\,\omega}
\end{equation}
of clean systems at zero temperature diverges \cite{vigh13} and
is then due to sliding of the hexagonal crystal as a whole
\cite{giamarchi05}. Its Drude weight $\:D=e^2n/m\:$ is determined
by carrier density $n$ and particle masses $m$, just as in the
absence of interactions \cite{kohn}; in this work we do not
consider crystalline disorder. In flat bands, where no kinetic
energy competes, we always expect a Wigner crystallized phase.
Conductance properties, quantified again by the Drude weight
$D$, cannot simply be proportional to $m^{-1}$ since $m$ is {\em
a priori\/} meaningless to parameterize kinetic energy
\cite{nomass}. In this work we determine $D$ of flat bands in
the presence of (static) Coulomb interactions. Our main result
is sketched in Fig.~2, below: $D$ exhibits an infinite number of
zeros at inverse band fillings, cf.\ Eq.~(\ref{commensurate}).
Near these zeros, the Drude weight rises linearly with slopes
that we estimate. Right in the middle between adjacent
insulating fillings phase transitions occur for which we expect
strongly enhanced conductivities.

\begin{figure}\vspace*{1cm}
\includegraphics[width=0.7\columnwidth]{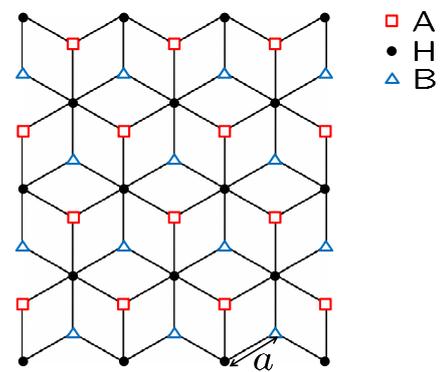}
\caption{Structure of the $\mathcal{T}_3$--lattice:
A ($\Box$) and B($\triangle$) sites are connected to
only 3 nearest H ({\protect\raisebox{-0.26ex}{\LARGE$\bullet$}})
sites, while conversely, H-sites are 6-fold coordinated.
The structure implies equal numbers of lattice sites,
$\:N_{\rm A}=N_{\rm H}=N_{\rm B}\:$.}
\end{figure}

We analyze flat bands in two dimensions exemplarily for the
$\mathcal{T}_3$--lattice \cite{doucot98}, which in literature
also is called Sutherland-- or dice--lattice, and which, in
principle, can be realized on the basis of cold quantum gases
\cite{bercioux09,moessner14}. It consists of three
different sublattices, (A,H,B), where A-- and B--sites are not
directly connected but only through H-sites, cf.\ Fig.~1.

Equal nearest neighbor hopping $t$ across bonds of lengths $a$
and zero hoppings between sites of larger distance gives rise
to energy bands $\varepsilon_k$ resembling much the band
structure of graphene with two Dirac points at the corners of
the hexagonal Brillouin zone and cone-like linearly dispersing
bands in their vicinity. Additionally, at zero energy, there is
a flat band extending throughout the Brillouin zone. In direct
generalization to graphene, the long wave length Hamiltonian
near one of the two inequivalent Dirac-points can be described
in spinor form
\begin{equation}\label{model}
H=\vf\:\bm{p}\cdot\bm{S}
\end{equation}
when using spin-1 operators $\:\bm{S}=(S_x,S_y)\:$
\cite{bercioux09,chamon10} (instead of spin-$\nicefrac{1}{2}$
operators as for graphene) to describe pseudo-spins that encode
now amplitudes on three sublattices (A,H,B), $\vf=3ta/\sqrt{2}$
being the Fermi velocity. The carrier momentum $\bm{p}$ in
(\ref{model}) formally serves as quantization axes for three
pseudo-spin eigenstates of energies
$\:\varepsilon_k^{(1,0,-1)}=(\vf |k|,0,-\vf |k|)\:$. Here, we
focus on the flat band $\varepsilon_k^{(0)}=0$ with
corresponding eigenstates $\:\psi_k^{(0)}=(-|k|{\rm e}^{-{\rm
i}\varphi},0,|k|{\rm e}^{{\rm i}\varphi})/\mathcal{N}_k\:$ that
manifest non-populated central H-sites ($|k|$ and
$\:\varphi=\arctan k_y/k_x\:$ denote, respectively, modulus and
direction of the carrier wave vector, and $\mathcal{N}_k$ a
normalization). A small spin-orbit term consisting of intrinsic
and Rashba contributions \cite{spinorbit} removes the Dirac point
by opening a gap which perturbatively renormalizes to increasing
values in the simultaneous presence of Coulomb interactions
\cite{spinorbit}; this isolates the flat band and, for small
spin-orbit coupling strength, leaves $\:\psi_k^{(0)}\:$
essentially unaffected.

Linear combinations within the $(N/3)$-fold degenerate space of
flat band Bloch states ($N=N_{\rm A}+N_{\rm H}+N_{\rm B}\:$
is the total number of lattice sites) allow to construct
strictly localized states around any H-site, such that the
entire amplitude resides just on one ring of six nearest A- and
B-sites surrounding that given H-site, of equal magnitude but of
alternating signs on A- and B-sites around the ring, and
vanishes elsewhere. This describes spatial localization of
`caged' quantum states in a strictly periodic lattice
\cite{sutherland}. At fillings $\:\nu=3N_{\rm e}^{(0)}/N\le
1\:$ of the flat band only $\nu N_{\rm H}$ out of all rings are
occupied, $N_{\rm e}^{(0)}$ being the flat band carrier number.

\begin{table}
\begin{tabular}{|lrrrr|}\hline
$1/\nu_{n_1n_2}$ & $n_1$ & $n_2$ & $\theta_{n_1n_2}$~~ & $\Delta_{n_1n_2}$\\ \hline
3 & 1 & 1 & 30. & 1\\
4 & 2 & 0 & 0. & 3\\
7 & 2 & 1 & 19.1066 & 2\\
9 & 3 & 0 & 0. & 3\\
12 & 2 & 2 & 30. & 1\\
13 & 3 & 1 & 13.8979 & 3\\
16 & 4 & 0 & 0. & 3\\
19 & 3 & 2 & 23.4132 & 2\\
21 & 4 & 1 & 10.8934 & 4\\
25 & 5 & 0 & 0. & 2\\
~~$\vdots$&&&&\\
1477 & 31 & 12 & 15.6886 & 6\\
1483 & 38 & 1 & 1.2886 & 5\\
1488 & 28 & 16 & 21.0517 & 1\\
1489 & 37 & 3 & 3.8606 & 3\\
1492 & 34 & 8 & 10.3327 & 5\\
1497 & 32 & 11 & 14.2536 & 4\\
1501 & 36 & 5 & 6.4171 & 15\\
1516 & 30 & 14 & 18.1432 & 3\\
1519 & 35 & 7 & 8.9483 & 2\\
1521 & 39 & 0 & 0. & 3\\
~~$\vdots$&&&&\\
\hline
\end{tabular}
\caption{Some values of inverse commensurate fillings
$\:\nu_{n_1n_2}^{-1}=n_1^2+n_2^2+n_1n_2\:$ satisfying
(\ref{commensurate}) where $(n_1,n_2)$ is one out of in
general 12 pairs of associated integers. The angles 
$\theta_{n_1n_2}=\arctan\sqrt{3}/(1+2n_1/n_2)$ in degrees
(chosen here as $0\le\theta_{n_1n_2}\le 30^\circ$) describe
how the corresponding Wigner crystal is oriented w.r.t.\
the $\mathcal{T}_3$-lattice. Differences separating adjacent
$\nu_{n_1n_2}^{-1}$ are called $\Delta_{n_1n_2}$.}
\end{table}

Long range Coulomb interactions $\:\sim
e^2/\kappa|\bm{r}-\bm{r'}|\:$ will lift the huge flat band
degeneracy by favoring maximum distances between carriers,
$\kappa$ is the dielectric constant. On a continuous space
the ground state would be a Wigner crystal \cite{meissner} of
lattice constant $\:b=\sqrt{3/\nu}\:a= \sqrt{2/\sqrt{3}n_{\rm
e}}\:$, where $a$ is the $\mathcal{T}_3$-lattice constant and
$n_{\rm e}=2/\sqrt{3}b^2$ the flat band carrier density. Here,
however, carriers cannot occupy arbitrary continuous positions
but must reside at H-sites (more precisely, on rings surrounding
H-sites) which constrains possible carrier positions to the
hexagonal H-lattice of lattice constant $\sqrt{3}a$ with which
$b$ is in general incommensurate at arbitrary $\nu$.
Commensurability between both hexagonal lattices necessitates
\begin{equation}\label{commensurate}
\frac{b}{\sqrt{3}a}=1/\sqrt{\nu_{n_1n_2}}\equiv
\sqrt{n_1^2+n_2^2+n_1n_2}\;,\quad n_1,n_2\in\mathbb{Z}
\end{equation}
for some pair of integers $(n_1,n_2)$. At fillings
$\nu_{n_1n_2}$ the Wigner crystal just ``fits'' to
$\mathcal{T}_3$-lattice sites. Tab.~1 lists some typical
examples for $\:\nu_{n_1n_2}^{-1}=n_1^2+n_2^2+n_1n_2\:$
satisfying (\ref{commensurate}). At commensurate fillings
$\nu_{n_1n_2}$ the conductivity vanishes, as demonstrated
below. We remark that seemingly random distances
$\:\sim\Delta_{n_1n_2}\:$ separate adjacent commensurate
fillings, with bigger $Delta_{n_1n_2}$ occurring rarer. This
resembles somewhat to prime numbers though, at the moment, we
are unable to estimate the asymptotic decay for the occurrence
of large $\Delta_{n_1n_2}$ as a function of the magnitude of
$\Delta_{n_1n_2}$.

Wigner crystallization spontaneously breaks translational and
rotational symmetries so that the ensuing crystal at
$\nu=\nu_{n_1n_2}$ will be oriented at angle
$0\le\theta_{n_1n_2}\le\pi/6$ w.r.t.\ the underlying
$\mathcal{T}_3$-lattice, where
$\:\tan\theta_{n_1n_2}=\sqrt{3}/(1+2n_1/n_2)=
\tan(\pi/3-\theta_{n_2n_1})\:$. Accounting additionally for the
trivial hexagonal symmetry,
$\:\theta_{n_1n_2}\to\theta_{n_1n_2}+n\pi/3\:$ for
$\:n=1,\ldots,5\:$, there are altogether 12 pairs of integers
$(n_1,n_2)$ characterizing commensurate Wigner lattices at same
flat band carrier density \cite{only6}. As a result, crystals
of adjacent commensurate densities will be more or less randomly
oriented, cf.\ Tab.~1.

How does the ground state look like at fillings
$\nu=\nu_{n_1n_2}+\delta\nu$ for $|\delta\nu|\ll\nu^2$ slightly
deviating from commensurate values $\nu_{n_1n_2}$? To
accommodate additional ($\delta\nu>0$) or missing
($\delta\nu<0$) carriers the system has basically two options:
{\em i)} develop two domains out of two adjacent commensurate
fillings each, or {\em ii)} accommodate, as $\nu_{n_1n_2}$-Wigner
crystal, $\delta\nu N_{\rm H}$ additional `electrons' at
interstitial places or leave $\delta\nu N_{\rm H}$ vacant Wigner
sites as `holes', depending on the sign of $\delta\nu$. Simple
energetic estimate supports the more homogeneous ground state
{\em ii)} near commensurate fillings since domains would suffer
energetically from additional charging, dipolar, and surface
contributions.

For conductance properties we analyze the current density
\begin{equation}\label{current}
\bm{j}=\vf\:\bm{S}\;,
\end{equation}
which for $\mathcal{T}_3$ resembles the relativistic expression
for the operator $\bm{j}$ of graphene \cite{vigh13,bercioux11}.
In $\mathcal{T}_3$, spin matrices
\begin{equation}\label{spinmat}
S_x=\frac{1}{\sqrt{2}}\begin{pmatrix}
0 & 1 & 0 \\
1 & 0 & 1 \\
0 & 1 & 0 \end{pmatrix}\;,\quad
S_y=\frac{1}{\sqrt{2}}\begin{pmatrix}
0 & -{\rm i} & 0 \\
{\rm i} & 0 & -{\rm i} \\
0 & {\rm i} & 0 \end{pmatrix}\;,
\end{equation}
substitute Pauli-matrices. We see that matrix elements
$\:\bra{\psi_k^{(0)}}\bm{j}\ket{\psi_{k'}^{(0)}}\:$ vanish
identically inside the flat band, as a result of vanishing
H-occupancy. In consequence, the system is insulating since the
conductivity is proportional to squares of these matrix
elements, in accordance with the above caging argument. Finite
conductivity necessitates finite amplitudes on H-sites which, in
turn, requires to destroy the destructive quantum interference
responsible for caging on hexagonal rings surrounding H-sites,
for example by derogating the precise balance in the occupation
probabilities of A- and B-sites through asymmetric electrostatic
forces induced by proximate charges.

For commensurate crystals the entire environs of each Wigner
site are hexagonally symmetric for any $(n_1,n_2)$. While
electrostatic forces created by vicinal Wigner sites will
act on all 6 A- and B-sites encircling a given Wigner site,
enhancing e.g.\ their on-site potential energies, and the
magnitudes of respective hoppings $t_{\rm A}$ and $t_{\rm B}$ to
nearest H-sites, no {\em difference\/} between A and B arises,
$\:t_{\rm A}=t_{\rm B}\:$, leaving thus caging intact.
Therefore, at fillings $\nu=\nu_{n_1n_2}$ the flat band stays
insulating in the presence of long range Coulomb interactions.

The situation can change at fillings
$\:\nu=\nu_{n_1n_2}+\delta\nu\:$. In the following we focus on
low flat band fillings, $\nu\ll 1$, such that $b\gg a$, cf.\
(\ref{commensurate}). While both types of `dopants', electrons
and holes, may distort the crystal in their vicinity
\cite{nodisorder} [however, at most by discrete displacements,
constrained to H-sites --- we do not consider elastic distortion
of the underlying $\mathcal{T}_3$-lattice here, having in mind
for example optical lattices] to relax local electrostatic
strain this effect will have minor impact on the conductivity,
and is therefore neglected in the following. Energetically,
interstitials prefer positions right in the middle of Wigner
triangles, at distances $b/\sqrt{3}$ to nearest Wigner-sites,
while vacancies are just unoccupied Wigner-sites.

Generally, electrons as well as each of the six Wigner sites next
to holes experience non-hexagonally symmetric environs which may
mobilize them. Crucial for emerging A-B-imbalance are
electrostatic forces acting {\em differently\/} on A- and
B-sublattices. Nearest electrostatic centers located on
perpendicular bisectors of the A- and B-sites still do {\em
not\/} achieve such an imbalance, albeit now $t_{\rm A,B}$'s may
vary in magnitude as we go around the 6 sites of an occupied ring
while mirror symmetry remains preserved. This kind of symmetry
occurs when $\:\theta_{n_1n_2}=\pi/6\:$ for $\delta\nu>0$
(electrons) and $\:\theta_{n_1n_2}=0,\ \pi/3\:$ for
$\delta\nu<0$ (holes). As a result, near fillings $\nu_{n_1n_2}$
where $\:\theta_{n_1n_2}=\{{\pi/6\atop 0\;\:\mbox{\footnotesize
or}\;\:\pi/3}\}\:$ we expect vanishing conductance yet over a
finite interval $\:\nu=\nu_{n_1n_2}\pm|\delta\nu|\:$.

At all other fillings $\:\nu=\nu_{n_1n_2}+\delta\nu\:$ (with
$\delta\nu\ne 0$) destructive quantum interference and local
topological localization of dopants is destroyed, H-sites
acquire non-zero amplitude, and a finite delocalizing kinetic
energy arises (which then has to compete with the Coulomb
energy as in the ordinary Wigner transition). In principle,
this allows dopants to spread across the entire
$\mathcal{T}_3$-lattice, resulting in finite
conductivity. To quantify the potential {\em difference\/}
between A- and B-sites of a hexagon cage due to an asymmetric
Coulomb field let us consider some charge at distance $d\gg a$
from the ring center which induces amplitude
\begin{eqnarray}\label{uh}
u_{\rm H}&\sim&\frac{e^2}{\kappa\vf}
\frac{a^2}{d^2}\Bigl({\textstyle\left(\frac{3}{2}\right)^{3/2}-
\frac{3}{\sqrt{2}}}\Bigr)\left|\left\{{\cos
3\theta_{n_1n_2}\atop\sin 3\theta_{n_1n_2}}\right\}\right|\\
&&{\rm for}\quad\left\{{{\rm electrons}\atop{\rm
holes}}\right\}\nonumber
\end{eqnarray}
at the ring center and at the same time on H-sites outside the
considered ring. It depends on the angle $\theta_{n_1n_2}$ of the
Wigner crystal orientation; $d=b/\sqrt{3}$ or $d=b$ for
interstitials (electrons) or vacancies (holes), respectively. The
dimensionless prefactor $\:e^2/\kappa\vf\:$ in (\ref{uh}) may be
regarded as `effective fine structure constant', parameterizing
the interaction strength \cite{fogler12}. Within a continuum
theory, employed from now on at $\nu\ll 1$ for transport
properties, the non-zero amplitude (\ref{uh}) on H-sites
effectively results in kinetic energy of dopants, associated
with a mass
\begin{equation}\label{meff}
m_{\rm eff}=1/(\vf a\,|u_{\rm H}|)\;,
\end{equation}
which promotes tunneling through Coulomb barriers to nearest
interstitial or vacancy places at distances $b/\sqrt{3}$ or $b$,
respectively.

As a result, dopants effectively experience a periodic Coulomb
potential landscape created by caged Wigner charges. Ignoring
correlations between dopants and using path integral methods,
the dopants matrix element
\begin{equation}\label{teh}
t_{\rm e/h}=\Delta{\rm e}^{-S_0}
\end{equation}
for hopping between adjacent minima can be estimated within a
dilute instanton gas approximation \cite{coleman}.
Eq.~(\ref{teh}) results from summing over all possible series
of independent instanton actions
\begin{equation}\label{s0}
S_0=2\sqrt{\frac{e^2m_{\rm eff}}{\kappa}}\int\limits_{-d/2}^{d/2}
\!\!\!\!{\rm d}x\;(x^2+3d^2/4)^{-1/4}=C_1\sqrt{\frac{e^2m_{\rm
eff}}{\kappa}} \sqrt{d}
\end{equation}
for motions in the inverted Coulomb potential
$\:-V(x)=-\frac{2e^2}{\kappa}/\sqrt{x^2+3d^2/4}\:$ by distances
$d$ during imaginary times. We assumed that dominant parts
of the Coulomb barrier are created by two charges at distances
$\sqrt{3}d/2$ `off the way' which holds true for electron as
well as for hole tunneling, $\:C_1\approx 2.096\:$ is a
numerical factor which can be expressed analytically in terms of
hypergeometric functions.

% C_1\approx 2.09580
% Hypergeometric2F1[-3/4, 1., 1/2, -1/3] -> 1.4760500783712442

Estimates for the ratio $\Delta$ of fluctuation determinants in
(\ref{teh}) are possible when considering the periodic Coulomb
potential $\:V_0(1-\cos 2\pi x/d)\:$ along the tunnel trajectory
with $\:V_0=\frac{2e^2}{\kappa d}(\frac{2}{\sqrt{3}}-1)\:$; then
\cite{zinnjustin}
\begin{equation}\label{delta}
\Delta=\frac{16V_0}{(m_{\rm eff}V_0)^{1/4}\sqrt{d/2}}\;.
\end{equation}

\begin{figure}\begin{center}
\includegraphics[width=0.9\columnwidth]{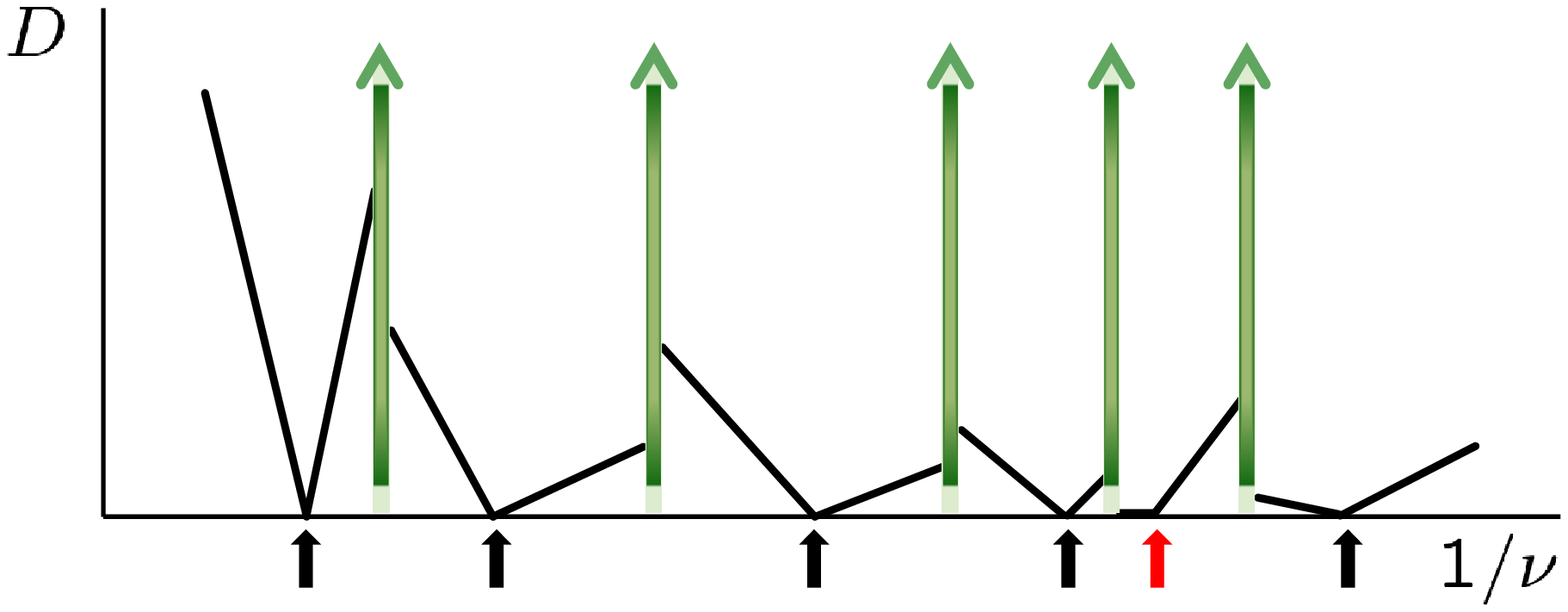}
\caption{Drude weight $D$ versus inverse flat band filling
$1/\nu$, schematically. When $\nu=\nu_{n_1n_2}$, cf.\
eq.~(\ref{commensurate}), $D=0$. Near these zeros (arrows),
$D$ rises linearly $\propto|\delta\nu|$ (unless
$\:\theta_{n_1n_2}=\pi/6+\mathbb{Z}\:\pi/3\:$ or
$\:\theta_{n_1n_2}=\mathbb{Z}\:\pi/3\:$ when $D$ stays
constantly zero to the left or right of $\nu_{n_1n_2}$,
respectively (red arrow)). In the middle between two adjacent
zeros of $D$, (green shaded rectangular areas), the system
exhibits a first order phase transition: there the Drude weight
is expected to rise considerably.}
\end{center}\end{figure}

Now we are in the position to estimate the Drude weight
$\:D=e^2n_{\rm D}/m_{\rm e/h}\:$ in eq.~(\ref{sigmaomega}).
Here, $\:n_{\rm D}=(|\delta\nu|/\nu)n_{\rm
e}=2|\delta\nu|/3^{3/2}a^2\:$ is the density and the band mass
for itinerant electron/hole type dopants takes values $\:m_{\rm
e/h}=1/(2t_{\rm e/h}d^2)\:$, respectively, cf.\ (\ref{teh})
[note that $m_{\rm e/h}$ differs from the kinetic energy mass
$m_{\rm eff}$]. Combining Eqs.\ (\ref{uh}---\ref{delta}), our
result for the Drude weight becomes
\begin{equation}\label{drudeweight}
D=\frac{e^2n_{\rm D}}{m_{\rm e/h}}=C_2{\textstyle\frac{e^2}{\kappa a}}
\gamma\nu^{-1/8}{\rm e}^{-S_0}\:\left|\left\{{\cos 3\theta_{n_1n_2}\atop\sin
3\theta_{n_1n_2}}\right\}\right|^{1/4}\:|\delta\nu|\;,
\end{equation}
% C_2 = 4 c2 / 3^{11/8} = 5.276083179349572;
% C_3 = 12 2^{3/4} c1 / \sqrt{2 - \sqrt{3}} = 20.427597699955857;
% c1 and c2 are constants from page XVII .
where
\begin{eqnarray*}
S_0&=&C_3(\nu\gamma)^{-3/2}\:\left|\left\{{\cos 3\theta_{n_1n_2}\atop\sin
3\theta_{n_1n_2}}\right\}\right|^{-1/2}\\
&&{\rm for}\;\left\{
{{\rm electrons}\atop{\rm holes}}\right\}\;,\;\gamma=
\left\{{3\atop 1}\right\}\;,
\end{eqnarray*}
and $\:C_2=5.276\:$ and $\:C_3=20.43\:$ being again
constants that can be expressed analytically.

Qualitatively, this result (\ref{drudeweight}) is sketched in
Fig.~2 versus $1/\nu$. Starting from zeros at commensurate
fillings $\nu_{n_1n_2}$ the Drude weight rises linearly with
small $|\delta\nu|=|\nu-\nu_{n_1n_2}|$. Slopes tend to decrease
with decreasing filling $\nu$, however, depending on the
orientations $\theta_{n_1n_2}$, they may differ strongly from
one zero to the next one. In symmetric cases,
$\:\theta_{n_1n_2}=\pi/6+\mathbb{Z}\:\pi/3\:$ (electrons) or
$\:\theta_{n_1n_2}=\mathbb{Z}\:\pi/3\:$ (holes) the Drude weight
stays constantly zero over a finite range of fillings on one
side of $1/\nu_{n_1n_2}$, cf.\ red arrow in Fig.~2. When the
density of electron type dopants from a lower density
$\nu_{n_1n_2}$ commensurate Wigner crystal equals the density of
hole type dopants from the next higher density $\nu_{n_1'n_2'}$
Wigner crystal, i.e.\ right in the middle between two adjacent
commensurate fillings $\nu_{n_1n_2}$ and $\nu_{n_1'n_2'}$ (green
shaded areas in Fig.~2), according to Landau's rule, a first
order solid to solid phase transition takes place. While
sweeping carrier density the system has to transform {\em all\/}
Wigner sites $(n_1,n_2)$ at crystal orientation
$\theta_{n_1n_2}$ into the next set of Wigner sites
$(n_1',n_2')$ at crystal orientation $\theta_{n_1'n_2'}$.
Therefore, at the phase transition caging should be destroyed
entirely and we expect a drastic increase of the Drude weight
during a transport measurement, up to values of the order
$\:e^2n_{\rm e}/m_{\rm eff}\:$ which exceeds the estimate
(\ref{drudeweight}) by a large factor
$\:\sim(\nu/|\delta\nu|){\rm e}^{S_0}\:$.

In conclusion, we have studied the Drude weight of flat band
insulators at zero temperature in the presence of long range
Coulomb interactions. A non-trivial function of density is
found, exhibiting infinitely many zeros at commensurate
fillings, given by eq.\ (\ref{commensurate}). Near those
fillings the Drude weight varies linearly with filling.
Sweeping the filling under transport conditions, first order
phase transitions of the electron crystal should cause
considerably enhanced Drude weights right in the middle between
adjacent commensurate fillings.

I am indebted to Reinhold Egger for many inspiring discussions
and acknowlegde gratefully sharing aspects of related research
with Dario Bercioux and with Daniel Urban. I thank Peter
Talkner for a valuable feed back on the manuscript, Peter
H\"anggi for providing fruitful working conditions, and the
State of Bavaria for still supporting fundamental research.


\begin{thebibliography}{99}

\bibitem{reviews} E.J. Bergholtz, Z. Liu, Int. J. Mod. Phys. B 
{\bf 27}, 1330017 (2013);
S.A. Parameswaran, R. Roy, S.L. Sondhi,
Comptes Rendus Physique {\bf 14}, 816 (2013),
and references therein.

\bibitem{neupert11} E. Tang, J.-W. Mei, X.-G. Wen,
Phys. Rev. Lett. {\bf 106}, 236802 (2011);
K. Sun, Z. Gu, H. Katsura, S. Das Sarma,
Phys. Rev. Lett. {\bf 106}, 236803 (2011);
T. Neupert, L. Santos, C. Chamon, C. Mudry,
Phys. Rev. Lett. {\bf 106}, 236804 (2011);
A.M. L"auchli, Z. Liu, E.J. Bergholtz, R. Moessner,
Phys. Rev. Lett. {\bf 111}, 126802 (2013).

\bibitem{daghofer12} S. Kourtis, J.W.F. Venderbos,
M. Daghofer, Phys. Rev. B {\bf 86}, 235118 (2012).

\bibitem{wang11} F. Wang, Y. Ran, Phys. Rev. B {\bf 84}, 241103(R) (2011);
A. Sterdyniak, C. Repellin, B.A. Bernevig,
N. Regnault, Phys. Rev. B {\bf 87}, 205137 (2013);
Z. Liu, E.J. Bergholtz, H. Fan, A.M.
L"auchli, Phys. Rev. Lett. {\bf 109}, 186805 (2012);
S. Kourtis, T. Neupert, C. Chamon, C. Mudry,
Phys. Rev. Lett. {\bf 112}, 126806 (2014).

\bibitem{nomass} Band flatness cannot be regarded as the limit
of band mass $m$ going to infinity, e.g.\ in a dispersion
$\:\varepsilon_k\propto k^2/2m\:$ which would yield finite
conductance in this limit due to the simultaneously increasing
density of states $\:\propto{\rm d}k(\varepsilon)/{\rm
d}\varepsilon\:$ that cancels the effect of decreasing velocity
$\:\propto{\rm d}\varepsilon_k/{\rm d}k\:$.

\bibitem{doucot98} J. Vidal, R. Mosseri, B. Dou\c{c}ot, Phys.
Rev. Lett. {\bf 81}, 5888 (1998).

\bibitem{chen14} L. Chen, T. Mazaheri, A. Seidel, X. Tang,
J. Phys. A: Math. Theor. {\bf 47}, 152001 (2014).

\bibitem{doucot00} J. Vidal, B. Dou\c{c}ot, R. Mosseri, P.
Butaud, Phys. Rev. Lett. {\bf 85}, 3906 (2000);
J. Vidal, P. Butaud, B. Dou\c{c}ot, R.
Mosseri, Phys. Rev. B {\bf 64}, 155306 (2001);
Z. Gul\'{a}csi, A. Kampf, D. Vollhardt,
Phys. Rev. Lett. {\bf 99}, 026404 (2007).

\bibitem{dassarma09} C. Wu, S. Das Sarma, Phys. Rev. B {\bf 77}, 235107 (2008);
S. Takayoshi, H. Katsura, N. Watanabe, H.
Aoki, Phys. Rev. A {\bf 88}, 063613 (2013).

\bibitem{stormer90} See, for example, H.W. Jiang, R.L. Willett,
H.L. Stormer, D.C. Tsui, L.N. Pfeiffer, K.W. West,
Phys. Rev. Lett. {\bf 65}, 633 (1990).

\bibitem{moraissmith09} C.-H. Zhang, Y.N. Joglekar, Phys. Rev. B
{\bf 75}, 245414 (2007);
O. Poplavskyy, M.O. Goerbig, C. Morais Smith, Phys. Rev. B {\bf 80},
195414 (2009).

\bibitem{tanatar} B. Tanatar, D.M. Ceperley, Phys. Rev. B
{\bf 39}, 5005 (1989).

\bibitem{vigh13} M. Vigh, L. Oroszl\'any, S. Vajna, P.
San-Jose, G. D\'avid, J. Cserti, B. D\'ora, Phys. Rev. B
{\bf 88}, 161413(R) (2013).

\bibitem{giamarchi05} R. Chitra, T. Giamarchi, Eur. Phys. J.
B {\bf 44}, 455 (2005); pinning of Wigner crystals by disorder
is beyond the scope of the present work.

\bibitem{kohn} W. Kohn, Phys. Rev. {\bf 123}, 1242 (1961).

\bibitem{bercioux09} D. Bercioux, D.F. Urban, H. Grabert, W. H\"ausler,
Phys. Rev. A {\bf 80}, 063603 (2009).

\bibitem{moessner14} For non time reversal invariant
generalizations cf.\ B. D\'ora, I.F. Herbut, R. Moessner,
arXiv:1402.6532, (2014).

\bibitem{chamon10} D. Green, L. Santos, C. Chamon, Phys.
Rev. B {\bf 82}, 075104 (2010);
B. D\'ora, J. Kailasvuori, R. Moessner, Phys. Rev. B
{\bf 84}, 195422 (2011).

\bibitem{spinorbit} C.L. Kane, E.J. Mele, Phys. Rev. Lett. {\bf 95},
226801 (2005).

\bibitem{sutherland} B. Sutherland, Phys. Rev. B {\bf 34}, 5208 (1986).

\bibitem{meissner} G. Meissner, H. Namaizawa, M. Voss, Phys. Rev. B
{\bf 13}, 1370 (1976); L. Bonsall, A.A. Maradudin, Phys. Rev. B
{\bf 15}, 1959 (1977).

\bibitem{only6} Unless $|n_1|=|n_2|$ or $n_i=0$ when only 6
different pairs $(n_1,n_2)$ exist.

\bibitem{bercioux11} D.F. Urban, D. Bercioux, M. Wimmer, W. H"ausler,
Phys. Rev. B {\bf 84}, 115136 (2011).

\bibitem{fogler12} In suspended graphene, for example, its
magnitude is of order $2.2/\kappa$, I. Sodemann, M. M.
Fogler, Phys. Rev. B {\bf 86}, 115408 (2012).

\bibitem{coleman} S. Coleman in {\em The Whys of Subnuclear
Physics}, Springer Subnuclear Series {\bf 15}, ed.\
by A. Zichichi (1979).

\bibitem{zinnjustin} J. Zinn-Justin, Nucl. Phys. B {\bf 192}, 125 (1981);
Z. Ambrozi\'{n}ski, Acta Physica Polonica B, {\bf 44}, 1261 (2013).

\end{thebibliography}
\end{document}